# First candidates for γ vibrational bands built on the [505]11/2⁻ neutron orbital in odd-A Dy isotopes

S. N. T. Majola[1, 2, 3]*, M. A. Sithole[3, 5], L. Mdletshe[2, 3], D. Hartley[8], J. Timár[9], B. M. Nyakó[9], J. M. Allmond[17], R. A. Bark[3], C. Beausang[18]‡, L. Bianco[12], T. D. Bucher[3, 31], S. P. Bvumbi[32], M.P. Carpenter[16], C. J. Chiara[19], N. Cooper[21], D. M. Cullen[4, 7], D. Curien[22], T. S. Dinoko[26], B. J. P. Gall[22], P. E. Garrett[12], P. T. Greenlees[4], J Hirvonen[4], U. Jakobsson[4], P. M. Jones[3], R. Julin[4], S. Juutinen[4], S. Ketelhut[4], B. V. Kheswa[1, 3], F.G. Kondev[16], A. Korichi[13], W.D. Kulp[20], T. Lauritsen[16], E. A. Lawrie[3, 5], L. Makhathini[3, 31], P. L. Masiteng[1], B. Maqabuka[3], E.A. McCutchan[16, 23], D. Miller[10, 27], S. Miller[6], A. Minkova[14], L. Msebi[3, 5], S. H. Mthembu[3, 5], J. Ndayishimye[3], P. Nieminen[4], P. Z. Ngcobo[29], S. S. Ntshangase[2], J. N. Orce[5], P. Peura[4], P. Rahkila[4], N. Redon[24], L. L. Riedinger[10], M. A. Riley[6], D. G. Roux[11], P. Ruotsalainen[4], J. Piot[22], J. Saren[4], J. F. Sharpey-Schafer[5], C. Scholey[4], O. Shirinda[3], J. Simpson[25], J. Sorri[15], I. Stefanescu[18], S. Stolze[4]†, J. Uusitalo[4], X. Wang[6, 28], V. Werner[21, 30], J.L. Wood[20], C.-H.Yu[17], S. Zhu[18] and G. Zimba[4].

[1]Department of Physics, University of Johannesburg, P.O. Box 524, Auckland Park 2006, South Africa
[2] University of Zululand, Department of Physics and Engineering, kwaDlangezwa, 3886, South Africa
[3]iThemba LABS, P. O. Box 722, Somerset–West 7129, South Africa
[4]University of Jyvaskyla, Department of Physics, P.O. Box 35, FI-40014, Finland
[5]University of the Western Cape, Department of Physics, P/B X17, Bellville 7535, South Africa
[6]Department of Physics, Florida State University, Tallahassee, FL 32306, USA
[7] Schuster Laboratory, University of Manchester, Manchester M13 9PL, United Kingdom.
[8]Department of Physics, U.S. Naval Academy, Annapolis, Maryland 21402, USA
[9]MTA Atomki, P.O. Box 51, H–4001 Debrecen, Hungary
[10]University of Tennessee, Department of Physics and Astronomy, Knoxville, Tennessee 37996, USA
[11]Department of Physics, Rhodes University, P.O. Box 94, Grahamstown 6140, South Africa
[12]University of Guelph, Department of Physics, Guelph, Ontario NIG 2WI, Canada
[13]CSNSM–IN2P3–CNRS, F–91405 Orsay Campus, France
[14]University of Sofia, Faculty of Physics, Sofia 1164, Bulgaria
[15]Sodankylä Geophysical Observatory, University of Oulu, Tähteläntie 62, FI-99600 Sodankylä
[16]Physics Division, Argonne National Laboratory, Argonne, Illinois 60439, USA
[17]Physics Division, Oak Ridge National Laboratory, Oak Ridge, TN 37831, USA
[18] Department of Physics, University of Richmond, Richmond, VA 23173, USA
[19]Nuclear Engineering Division, Argonne National Laboratory, Argonne, IL 60439, USA
[20]School of Physics, Georgia Institute of Technology, Atlanta, GA 30332, USA
[21] Wright Nuclear Structure Laboratory, Yale University. New Haven, CT 06520, USA
[22]Université de Strasbourg, CNRS, IPHC, UMR 7178, F-67000 Strasbourg, France
[23] National Nuclear Data Centre, Brookhaven National Laboratory, Upton, NY 11973, USA
[24] Institut de Physique Nucléaire Lyon, IN2P3-CNRS, Lyon, F-69622 Villeurbanne, France
[25] STFC Daresbury Laboratory, Daresbury, Warrington, WA4 4AD, UK
[26] National Metrology Institute of South Africa, P/B X34, Lynnwood Ridge, Pretoria 0040, South Africa
[27] TRIUMF, Vancouver, British Columbia, V6T2A3C, Canada
[28]Department of Physics, California Polytechnic State University, San Luis Obispo, CA 93407, USA
[29]Department of Physics, University of Cape Town, P/B X3, Rondebosch 7701, South Africa
[30] Institut für Kernphysik, Technische Universität Darmstadt, Schlossgartenstr. 9, 64289 Darmstadt, Germany
[31]Department of Physics, Stellenbosch University, P/B X1, Matieland 7602, South Africa
[32]National Radioactive Waste Disposal Institute, NECSA, Pelendaba, South Africa
Present addresses: † Physics Division, Argonne National Laboratory, Argonne, Illinois 60439, USA
‡ Deceased
*email address; smajola@uj.ac.za

## Abstract

Rotational structures have been measured using the Jurogam II and GAMMASPHERE arrays at low spin following the $^{155}$Gd($\alpha$,2n)$^{157}$Dy and $^{148}$Nd($^{12}$C, 5n)$^{155}$Dy reactions at 25 and 65 MeV, respectively. We report high-*K* bands, which are conjectured to be the first candidates of a $K^{\pi}= 2^{+}$ γ vibrational band, built on the [505]11/2$^{-}$ neutron orbital, in both odd-A $^{155, 157}$Dy isotopes. The coupling of the first excited $K=0^{+}$ states or the so-called β vibrational bands at 661 and 676 keV in $^{154}$Dy and $^{156}$Dy to the [505]11/2$^{-}$ orbital, to produce a $K^{\pi}=11/2^{-}$ band, was not observed in both $^{155}$Dy and $^{157}$Dy, respectively. The implication of these findings on the interpretation of the first excited 0$^{+}$ states in the core nuclei $^{154}$Dy and $^{156}$Dy are also discussed.

## I. Introduction

The conventional interpretation of first excited $K^{\pi}=2^{+}$ bands, that consistently appear in every even-even deformed nucleus, has been that they are a shape vibration in the γ degree of freedom perpendicular to the symmetry axis [1, 2, 3]. These bands have both odd- and even-spin states. The energy staggering between level energies of the even- and odd-spin members of the γ-band provides an insight into the nature of triaxiality in the nucleus [4]. Furthermore, since $\Delta K = 2$ for the out-of-band γ-rays decaying out of both bands, M1 components are *K*-forbidden in the $\Delta I =0$ (from the even-spin members to the ground band) and $\Delta I = \pm 1$ (from the odd-spin members to the ground band) transitions [5, 6]. Similarly, due to the very inhibited M1 components, the $\Delta I=1$ transitions between the even- and odd-spin members of the $K^{\pi}=2^{+}$ bands, particularly in the 150 to 170 mass region, are observed to be very weak at most [5–9].

In the A≈160 region, where the focus of this work is, the $K^{\pi}=2^+$ bands usually lie about 1.0 MeV above the yrast line, which makes it difficult to populate such states in heavy-ion fusion-evaporation reactions as they are embedded in other structures which compete for intensity. Consequently, it is challenging to find cases where these bands have been identified beyond spin $14^+$. However, wherever they are observed in the transitional rare-earth region, one of their notable features is that they, particularly the odd-spin members, faithfully track their intrinsic configuration (usually the ground state band in even-even nuclei), as a function of spin [7–14].

In odd-A nuclei, the single nucleon in a Nilsson orbital $[N,n_x,\Lambda]\Omega$ are expected to couple to the collective states of the even-even core. For instance, a single-particle orbital with a Nilsson quantum number $\Omega$, would couple to the so-called β vibrational band of the even-even core to form a $K=\Omega$ band in the neighbouring odd-N nucleus. Similarly, the coupling of the single-particle state(s) with the $K^{\pi}=2^+$ excitations may result into either a parallel mode or an antiparallel mode which give $K_> = (\Omega+2)$ or $K_< = (\Omega-2)$, respectively. In the transitional rear earth region, examples whereby bands built on both $K_<$ and $K_>$ couplings have been reported remain a rare occurrence [15, 16].

In recent years, there have been a few cases of in-beam (heavy-ion reaction) studies which have reported rotational bands built on a state resulting from the $K_>$ coupling [15, 17, 18, 55]. Failure to observe the $K_>$ band may also be attributed to the fact that these bands are typically situated ≈ 1 MeV above the yrast line in most nuclei (particularly in the A≈160 region) and thus are most likely to be weakly populated in most fusion-evaporation reactions.

In this present paper we report and investigate the identification of structures built on the $[505]11/2^-$ neutron orbital in odd-A $^{155}$Dy and $^{157}$Dy, which have been identified as candidates for the $K^{\pi}=2^+$ γ-bands coupled to the $[505]11/2^-$ neutron orbital. It is worth noting that this work focuses mainly on results relating to bands built on the $h_{11/2}$ orbital. Results pertaining to bands built on other prominent single-particle orbitals close to the Fermi surface in N ≈ 90 nuclei such as $i_{13/2}$, $h_{9/2}$ and $f_{7/2}$, particularly for $^{157}$Dy, have been presented in a separate publication [19]. A previous data set has also been examined in order to identify levels built on top of the $[505]11/2^-$ bands in $^{155}$Dy [13, 20, 21].

## II. Experimental details and analysis

Excited states of $^{157}$Dy were populated using the $^{155}$Gd (α, 2n) reaction at a beam energy of 25 MeV. The $^{155}$Gd target was 0.98 mg/cm$^2$ thick with a purity of 91%. Gamma–rays following the reaction were detected with the JUROGAM II multi–detector array [22] in JYFL, Jyväskylä, Finland. This γ–ray array consists of 39 HPGe detectors, each in its own BGO shield. Almost 2/3 of the array (24 detectors) comprises of EUROGAM II clover detectors [23] while the rest are coaxial EUROGAM Phase 1 and GASP type detectors [24]. Approximately 14 x 10$^9$ γ-γ events were unfolded from the data in the off-line analysis and replayed into a γ-γ Radware [25] matrix from which the level scheme for the nucleus of interest was created.

The spins and parities for the new rotational structures were assigned using the Directional Correlation from Oriented States (DCO ratios/$R_{DCO}$) [26, 27] and linear polarization anisotropy ($A_p$) methods [28]. While $A_p$ are defined in [29], the DCO matrices were constructed using data from detectors in the rings at 158° and 86°+94°. Thus the $R_{DCO}$ for the JUROGAM II array in this work is;

$$R_{DCO} = I_{\gamma 1}(at\ 158^o : gated\ on\ \gamma_2\ near\ 90^o) / I_{\gamma 1}(near\ 90^o : gated\ on\ \gamma_2\ at\ 158^o) \quad (1)$$

When a gate is placed upon a known stretched quadrupole transition, an $R_{DCO}$ ratio close to 0.5 and 1.0 is expected for a stretched dipole and quadrupole transitions, respectively. The magnetic and electric nature of the transitions was inferred from their $A_p$ values. The $A_p$ values yield $A_p > 0$ and $A_p < 0$ for stretched electric and magnetic transitions, respectively.

Finally, to identify levels built on the [505]11/2$^-$ bands in $^{155}$Dy, previous data sets were examined. These data resulted from the $^{148}$Nd ($^{12}$C, 5n) $^{155}$Dy [13, 20] and $^{124}$Sn($^{36}$S,5n) $^{155}$Dy reactions [21], which were both collected using the GAMMASHERE array. Details about the experimental setup and analysis of these experiments can be found in Refs. [13, 20, 21].

# III. Level schemes
## A. $^{155}$Dy: Bands 1 and 2

Fig. 1 shows a partial level scheme of $^{155}$Dy deduced from this work. The rotational band built on the 234-keV level shown in Fig. 1 is the well-known high-$K$ band from the comprehensive work of T. Brown *et al.,* [30] and R. Vlastou *et al.,* [31]. This structure is understood to arise from the [505]11/2$^-$ neutron configuration and is interlinked by strong M1 transitions [30, 31]. As can be seen in Fig. 1, the present study has not only confirmed the validity of excitation levels associated with the [505]11/2$^-$ orbital (up to 31/2$^-$) but has also established new structures built on them. The new levels and γ-rays of $^{155}$Dy, established from our work, are labelled in red and identified as Bands 1 and 2, while structures known from previous studies are in black. Measured properties of γ–rays and rotational levels observed in this work are listed in Table 3. The spectrum in Fig. 3 (a), shows some of the new intra-band as well as inter-band transitions decaying out of the newly established structures.

DCO measurements could only be carried out for the 830-keV transition depopulating the 1064-keV level. The DCO ratio for this transition is consistent with it being a stretched E2 transition. When taking the spin-parity selection rules into account the possible spin and parity of the 1064-keV level becomes 11/2$^+$ or 15/2$^-$. Assuming that Bands 1 and 2 form a high-$K$ band structure connected by M1 transitions, we conclude that the 11/2$^+$ assignment is unlikely as it would mean that the 1305-, 1560- and 1827-keV levels decay to the [505]11/2$^-$ band via hindered M2 transitions through the 241, 410 and 409-keV γ-rays, respectively. For this reason, the $E_x$ = 1064-, 1305-, 1560-, 1827-, 2104-, 2390-, 2699- and 2963-keV levels of Bands 1 and 2 are tentatively assigned to have spin-parity of 15/2$^-$, 17/2$^-$, 19/2$^-$, 21/2$^-$, 23/2$^-$, 25/2$^-$, 27/2$^-$ and 29/2$^-$, respectively.

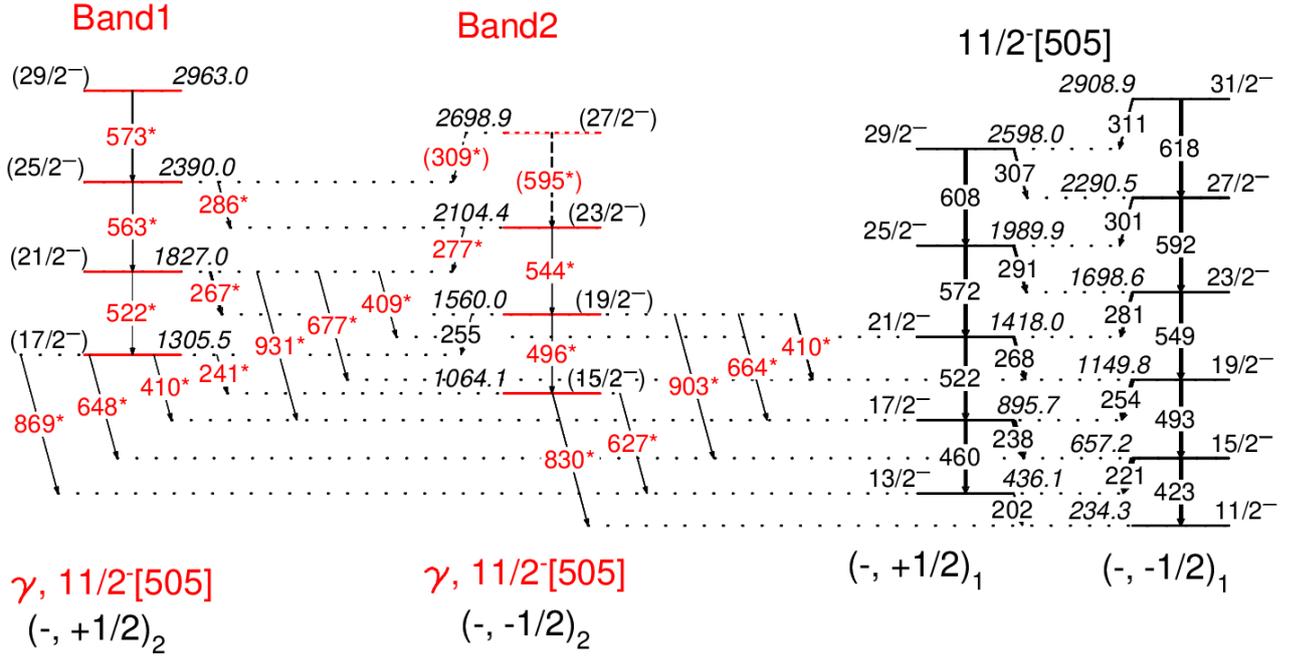

*Figure 1:* (Color online) Partial level scheme of $^{155}$Dy deduced from the current work. It shows band structures built on the [505]11/2$^-$ neutron orbital in $^{155}$Dy. New rotational structures, established in this work that have been identified as the γ band built on [505]11/2$^-$ bands are labeled in red (and marked with asterisk symbols) while known bands, deduced from previous in-beam works, are labeled in black.

### B. $^{157}$Dy: Bands 3 and 4

The level scheme of $^{157}$Dy is well known from the previous in-beam work of Hayakawa et al., [53], Pipidis [41] et al., and Riley [52] et al., which comprehensively explored the structural behaviour of this nucleus at medium to high spin. These studies deduced that the level scheme of this nucleus is divided into three parts with each band structure based on the [521]3/2$^-$, [651]3/2$^-$ and [505]11/2$^-$ orbitals. The latter is the main focus of this work, thus Fig. 2 only shows a partial level scheme of $^{157}$Dy, which contains the [505]11/2$^-$ bands as well as the new structures built on them. The rest of the level scheme of $^{157}$Dy, based on the current experimental results, has been published elsewhere [19].

The rotational structures shown in Fig. 2 form a high-$K$ band built on an isomeric state at 199-keV with a half-life of 21.6 ms. Previous studies [41, 53] assigned these structures to the [505]11/2⁻ Nilsson orbital and reported them up to spins 75/2⁻ and 69/2⁻, respectively [41]. This work has managed to confirm and validate the placements made by previous studies up to 29/2⁻ and 31/2⁻. As can be seen in Table 4, the DCO and polarization values deduced from this work are also in agreement with the spin and parity assignments from previous work [32, 41, 53].

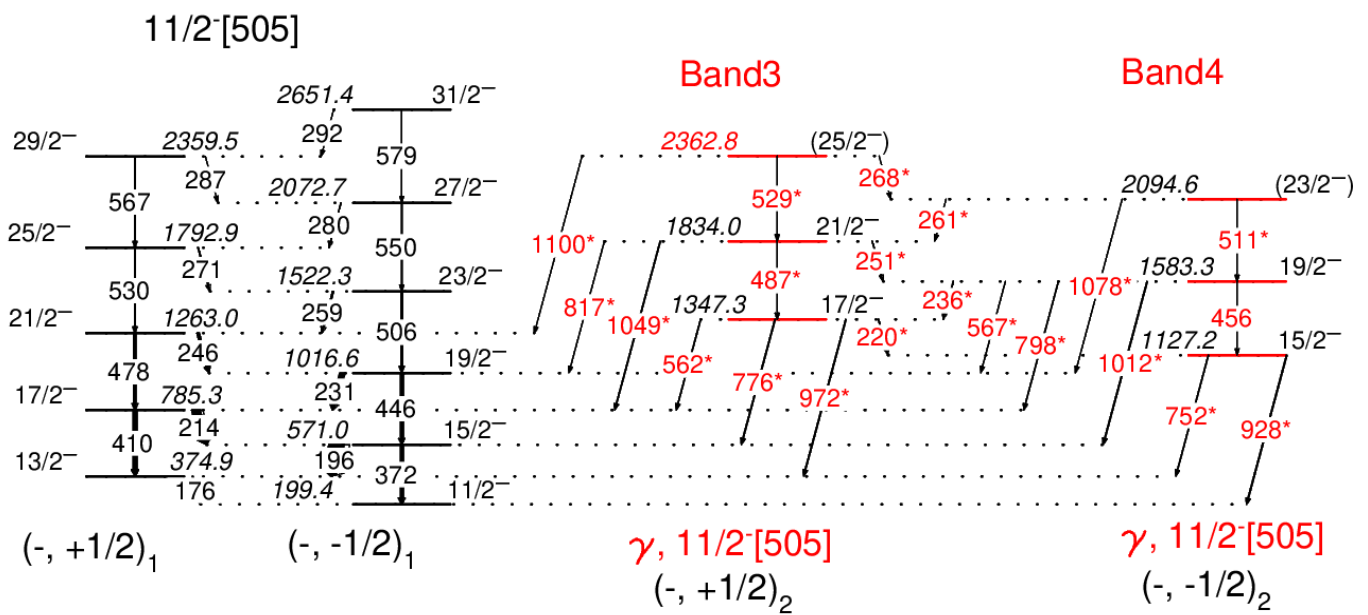

*Figure 2: (Color online) Partial level scheme of ¹⁵⁷Dy deduced from the current work. It shows band structures built on the [505]11/2⁻ neutron orbital in ¹⁵⁷Dy. New rotational structures, established in this work that have been identified as the γ band built on [505]11/2⁻ bands are labeled in red (and marked with asterisk symbols) while known bands, deduced from previous in-beam works, are labeled in black.*

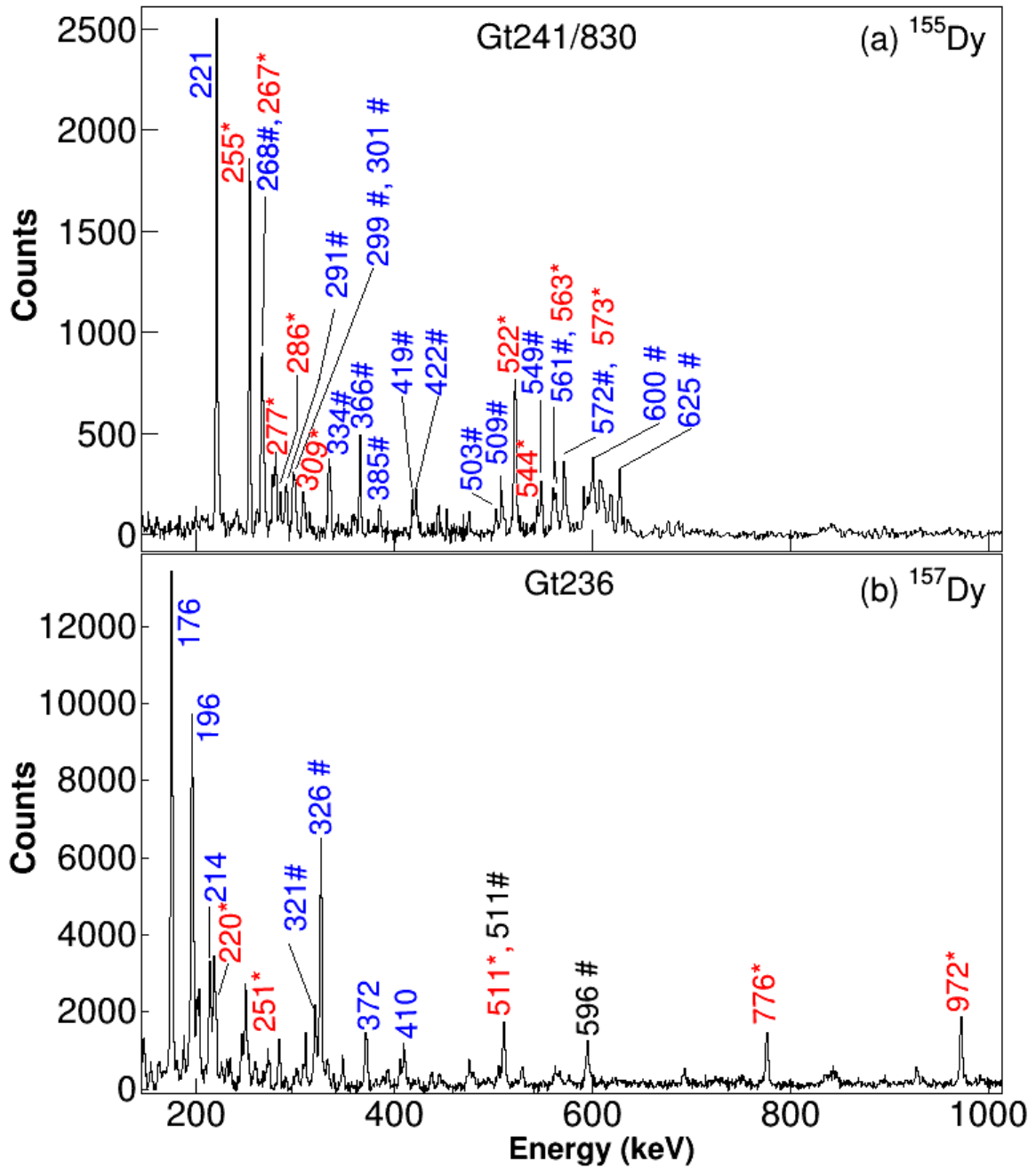

*Figure 3:* (Color online): (a) Double and (b) Single gated coincidence spectra for the γ-band built on the [505]11/2⁻ neutron orbital in $^{155}$Dy and $^{157}$Dy, respectively. Transitions corresponding to these newly found structures are labeled in red (and marked with asterisk symbols) while contaminants from other reaction channels and/or other bands of $^{155,\ 157}$Dy, not associated with the cascade of interest are denoted by hash (#) symbols. Previously known transitions in the nucleus and/or decay path of interest are also highlighted in blue but are unmarked.

Fig. 2 shows that this study has also identified a new strongly coupled band structure that decays to the [505]11/2⁻. New bands are identified as Bands 3 and 4, and both consist of three rotational levels, connected to each other by relatively strong inter-linking transitions compared to their weak intra-band transitions. As can be seen in Fig. 2, these newly found rotational levels decay solely to the [505]11/2⁻ bands through medium to high energy transitions. The DCO and polarization measurements carried out for the 928-keV transition depopulating the 1127-keV (depopulating Band 4) level are consistent with it being a pure stretched E2. The most probable spin and parity assignment for the 1127-keV level then becomes 15/2⁻. Similarly, for the 1347-keV level in Band 3, the DCO and polarization measurements for the transition decaying out of this level are consistent with it being a 17/2⁻. Assuming that Bands 3 and 4 are signature partners (based on the 1127-keV) connected by interlinking M1 transitions, the $E_x$ = 1127-, 1347-, 1583-, 1834-, 2095-, and 2363-keV levels of Bands 3 and 4 are then respectively assigned to have spin-parity of 15/2⁻, 17/2⁻, 19/2⁻, 21/2⁻, 23/2⁻ and 25/2⁻. Polarization and DCO measurements were also carried out for some of the inter-band transitions decaying out of Bands 3 and 4 where possible. These measurements are also consistent with the proposed spin and parity assignments. Fig. 3(b) presents a gated spectrum showing γ-rays associated with Bands 3 and 4.

## IV. Discussion

In this work we first attempt to explain and/or interpret the rotational behaviour of the new bands using the Cranked Shell Model (CSM) [44, 51]. Fig. 4 shows a CSM calculation of the Routhians e' of $^{157}$Dy, deduced using a modified oscillator potential. Concerning the interpretation of the structures in $^{155}$Dy, we rely on the CSM diagram of $^{155}$Dy given in Fig. 3.6 of Ref. [30]. Lastly, the labelling of the quasi–neutron states shown in Fig. 4, are given in Table 1.

## A. Bands 1, 2, 3 and 4

As can be seen in Figs. 1 and 2, the properties of these four new bands appear to share a lot of similarities. In effect, when comparing the structural behaviour of the new rotational states in $^{155}$Dy namely, Bands 1 and 2, with those of Bands 3 and 4 in $^{157}$Dy, it is clear that we are dealing with the same structure. First of all, in both $^{155}$Dy and $^{157}$Dy, the new bands form high-$K$ band structures (with $K^{\pi} = 15/2^{-}$), which decay solely to the [505]11/2$^{-}$ bands. The bandhead energies of these high-$K$ structures start at about 900 keV relative to the [505]11/2$^{-}$ bands and have a similar decay pattern. One of the striking features about these $K^{\pi} = 15/2^{-}$ bands is that in both $^{155}$Dy and $^{157}$Dy the energies of the interlinking M1 ($\Delta I = 1$) transitions between them are comparable to those connecting rotational levels of the $K = 11/2^{-}$ bands. This is clearly demonstrated in Table 2 where the energy difference (per given spin) between these M1 transitions averages about 3.5- and 6-keV for both $^{155}$Dy and $^{157}$Dy.

Furthermore, the rigid rotor and Routhian plots shown in Figs. 5 and 6 (c, d) further demonstrate similarities between the structures in question. In fact, the newly established high-$K$ structures in both $^{155}$Dy and $^{157}$Dy respectively run parallel to the [505]11/2$^{-}$ bands, as a function of spin. The aligned angular momenta of these high-$K$ structures are also plotted in Fig. 6 (a) and (b). In both cases, just like the [505]11/2$^{-}$ bands, the $K^{\pi} = 15/2^{-}$ bands appear to have a gradual upbend, but are consistently, at the most, half a unit above the [505]11/2$^{-}$ bands.

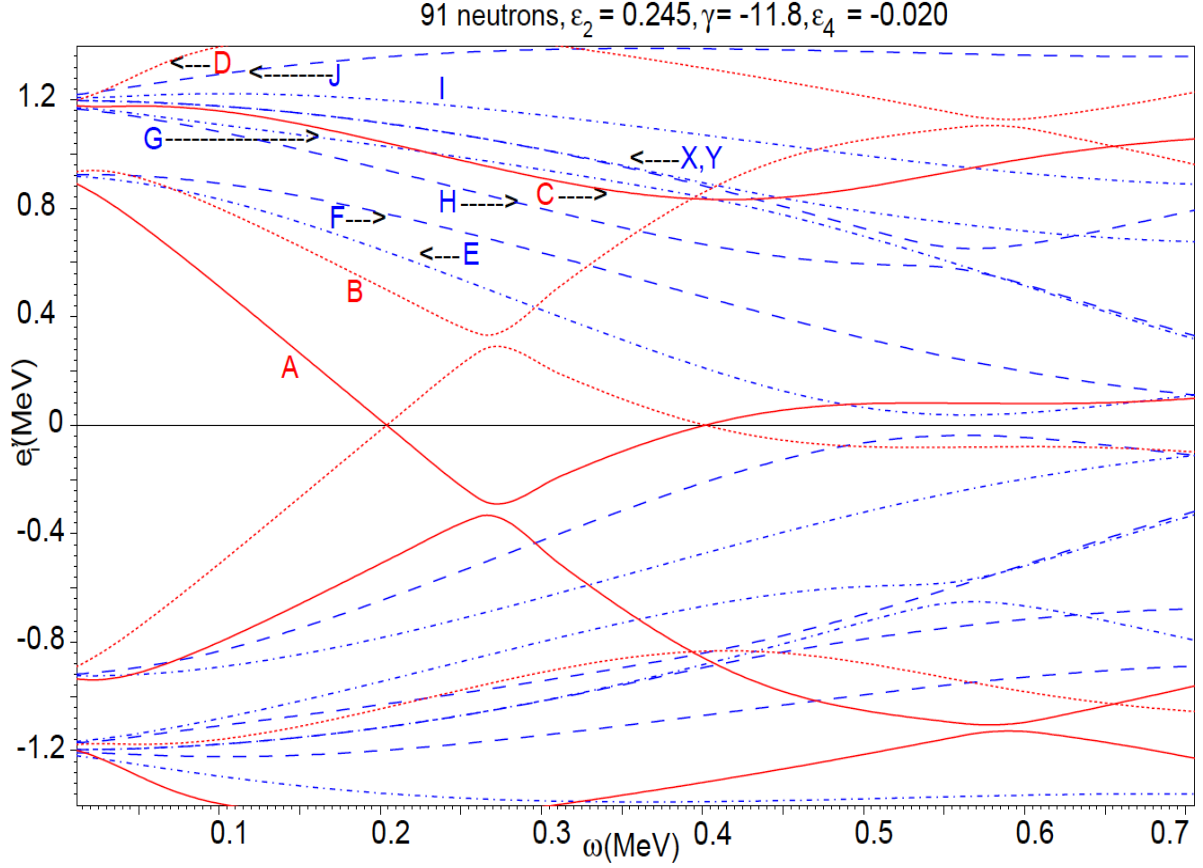

**Figure 4:** *(Color online) Neutron quasi–particle Routhians e', as a function of $\hbar\omega$, calculated for $^{157}$Dy using CSM with parameters $\varepsilon_2 = 0.245$, $\varepsilon_4 = -0.020$ and $\gamma = -11.8°$. The solid and dotted lines colored in red represent positive–parity states with signatures $\alpha = +1/2$ and $-1/2$, respectively. The dot–dashed and dashed lines colored in blue represent negative–parity states with $\alpha = +1/2$ and $-1/2$, respectively. The corresponding quasi–neutron labeling is given in Table 1.*

A plot for the *B(M1)/B(E2)* ratios deduced for the newly established bands (high-*K* bands built on the [505]11/2$^-$ structures) is shown in Fig. 7. The ratios in $^{155}$Dy show further similarities with those measured in $^{157}$Dy. In effect, in both cases, ratios of both structures exhibit the same trend. With these pieces of information, we therefore conclude that we are dealing with the same structure in both $^{155}$Dy and $^{157}$Dy.

*Table 1: Labelling of the quasi–neutron states in Fig. 4, as well as that of the proposed configuration assignments for new band structures in $^{155}$Dy and $^{157}$Dy, where $(\pi, \alpha)_n$ represent the $n^{th}$ rotational sequence with signature α and parity π.*

| band label | Signature | Label of quasi-neutron states | Configuration assignment |
|---|---|---|---|
| $^{155}$Dy | | | |
| Band 1 | α = + 1/2 | – | $\gamma \otimes h_{11/2}[505]11/2^-$ |
| Band 2 | α = - 1/2 | – | $\gamma \otimes h_{11/2}[505]11/2^-$ |
| (–,–1/2)$_1$ | α = - 1/2 | – | $h_{11/2}[505]11/2^-$ |
| (–,+1/2)$_1$ | α = + 1/2 | – | $h_{11/2}[505]11/2^-$ |
| $^{157}$Dy | | | |
| – | α = + 1/2 | A | $i_{13/2}[651]3/2^+$ |
| – | α = - 1/2 | B | $i_{13/2}[651]3/2^+$ |
| – | α = + 1/2 | C | $i_{13/2}[660]1/2^+$ |
| – | α = - 1/2 | D | $i_{13/2}[660]1/2^+$ |
| – | α = + 1/2 | E | $h_{9/2}[521]3/2^-$ |
| – | α = - 1/2 | F | $h_{9/2}[521]3/2^-$ |
| – | α = - 1/2 | G | $f_{7/2}[523]5/2^-$ |
| – | α = + 1/2 | H | $f_{7/2}[523]5/2^-$ |
| – | α = + 1/2 | I | $h_{9/2}[530]1/2^-$ |
| – | α = - 1/2 | J | $h_{9/2}[530]1/2^-$ |
| (–,–1/2)$_1$ | α = - 1/2 | Y | $h_{11/2}[505]11/2^-$ |
| (–,+1/2)$_1$ | α = + 1/2 | X | $h_{11/2}[505]11/2^-$ |
| Band 3 | α = + 1/2 | – | $\gamma \otimes h_{11/2}[505]11/2^-$ |
| Band 4 | α = - 1/2 | – | $\gamma \otimes h_{11/2}[505]11/2^-$ |

**Table 2:** *Comparisons of M1 transitions (in keV) in the $K^\pi = 11/2^-$ and $K^\pi = 15/2^-$ bands, per given spin, of $^{155}$Dy and $^{157}$Dy isotopes, respectively.*

|  | 155Dy | | 157Dy | |
| --- | --- | --- | --- | --- |
| Spin(ℏ) | $K = 11/2^-$ | $K = 15/2^-$ | $K = 11/2^-$ | $K = 15/2^-$ |
| $17/2^-$ | 238 | 241 | 214 | 220 |
| $19/2^-$ | 255 | 254 | 231 | 236 |
| $21/2^-$ | 267 | 268 | 246 | 251 |
| $23/2^-$ | 277 | 281 | 259 | 261 |
| $25/2^-$ | 286 | 291 | 271 | 268 |

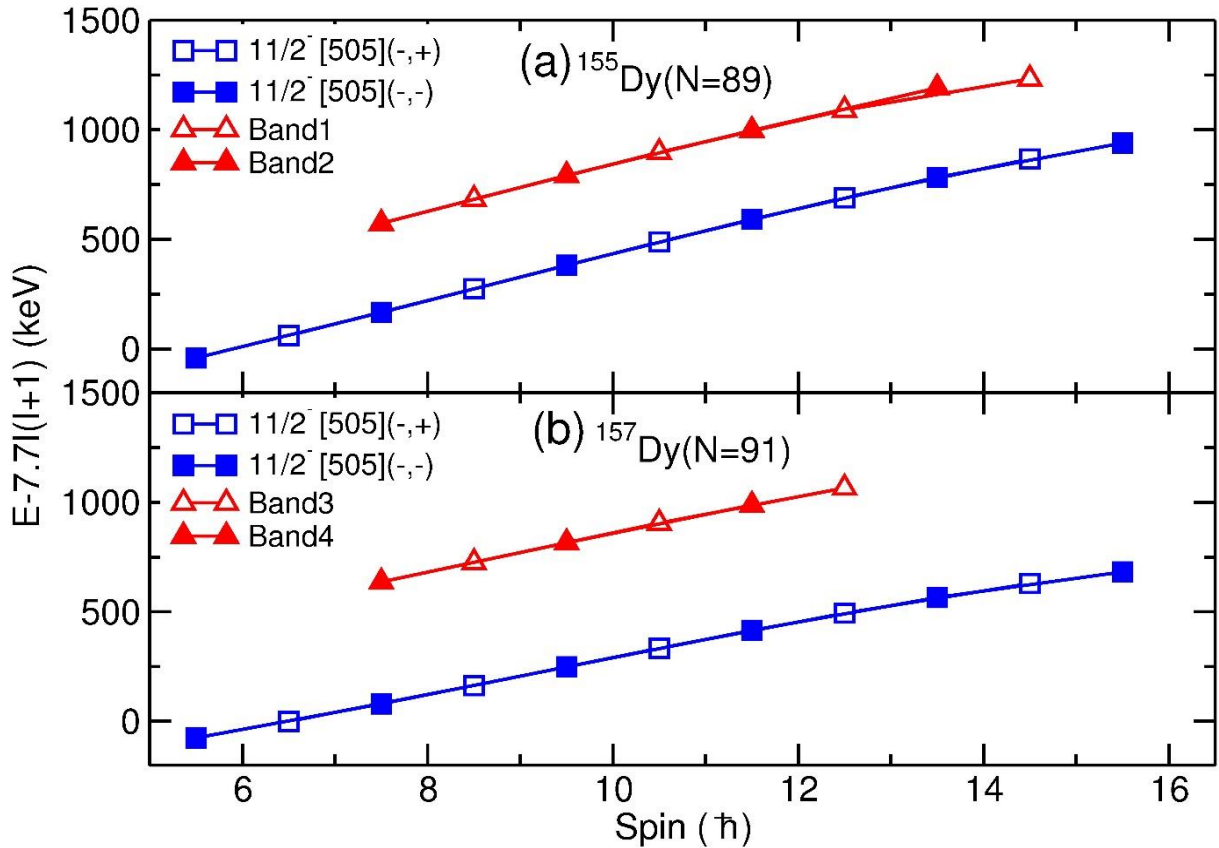

**Figure 5:** *(Color online) Plots of excitation energy minus a rigid rotor energy for bands observed in (a) $^{155}$Dy and (b) $^{157}$Dy. Open symbols have α = +1/2 while closed symbols have α = -1/2.*

In a quest to identify the microscopic nature of these $K^{\pi} = 15/2^-$ bands, we look at the possible ways through which they can arise or be formed. The band-head energies of these structures are about 1 MeV, thus suggesting that they could either be a single-particle orbital or a three-quasi-particle excitation or a single-phonon γ vibrational band. In the Nilsson model, there are no suitable single-neutron orbital in the Fermi surface which could give $K^{\pi} = 15/2^-$. This leaves the three-quasi-particle and γ vibration arguments as the only possibility. However, the Cranked Shell Model (see Fig. 4 and Fig. 3.6 of Ref. [30] for both $^{157}$Dy and $^{155}$Dy, respectively) also does not predict any plausible single-neutron orbital combinations that can form a three-quasi-particle configuration suitable to explain the features of the bands in question. This rules out the possibility of describing these bands as a three-quasi-particle configuration. It is therefore likely that the $K^{\pi} = 15/2^-$ bands in both $^{155}$Dy and $^{157}$Dy can best be described as the $K^{\pi} = 2^+$ single-phonon γ vibrational bands coupled to the [505]11/2$^-$ orbital.

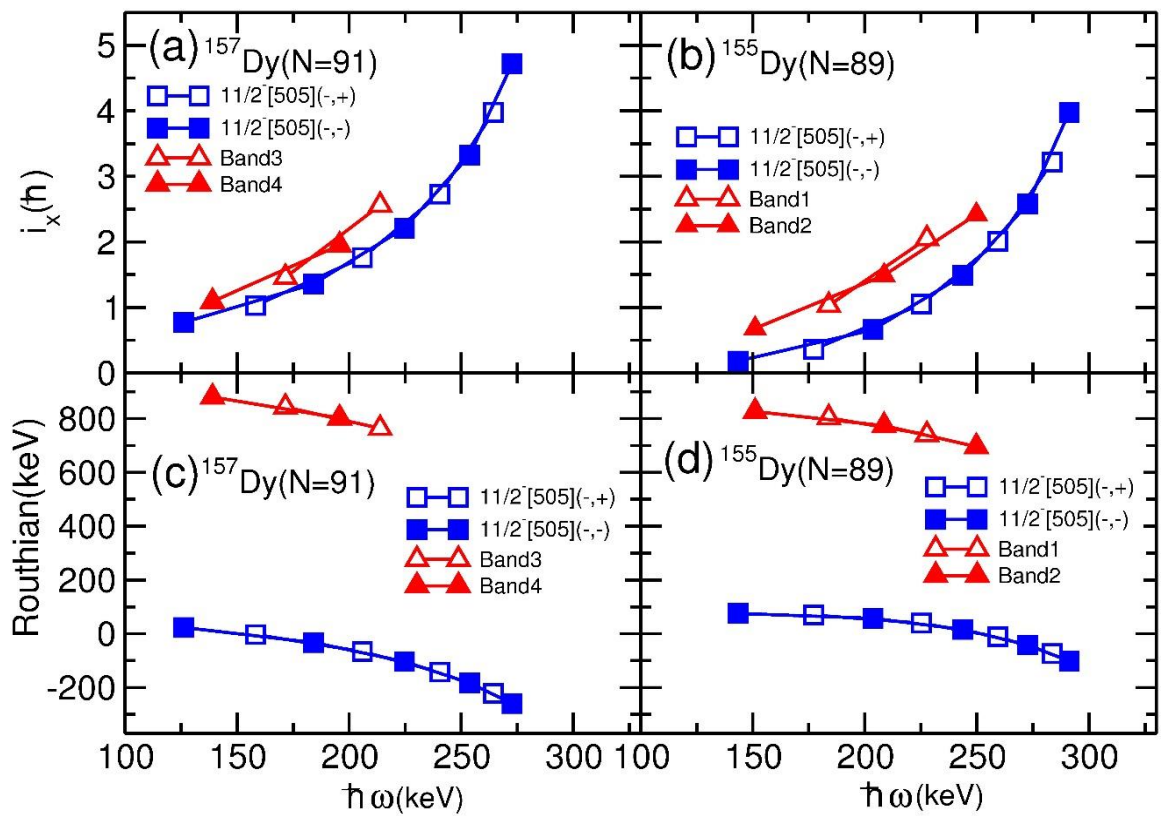

*Figure 6:* (Color online) Plots of alignment $i_x$ and Routhians e' for bands in $^{155}$Dy and $^{157}$Dy. Harris parameters used are $J_0 = 32\ \hbar^2\ MeV^{-1}$ and $J_1 = 34\ \hbar^4\ MeV^{-3}$. Open symbols have α = +1/2 while closed symbols have α = -1/2.

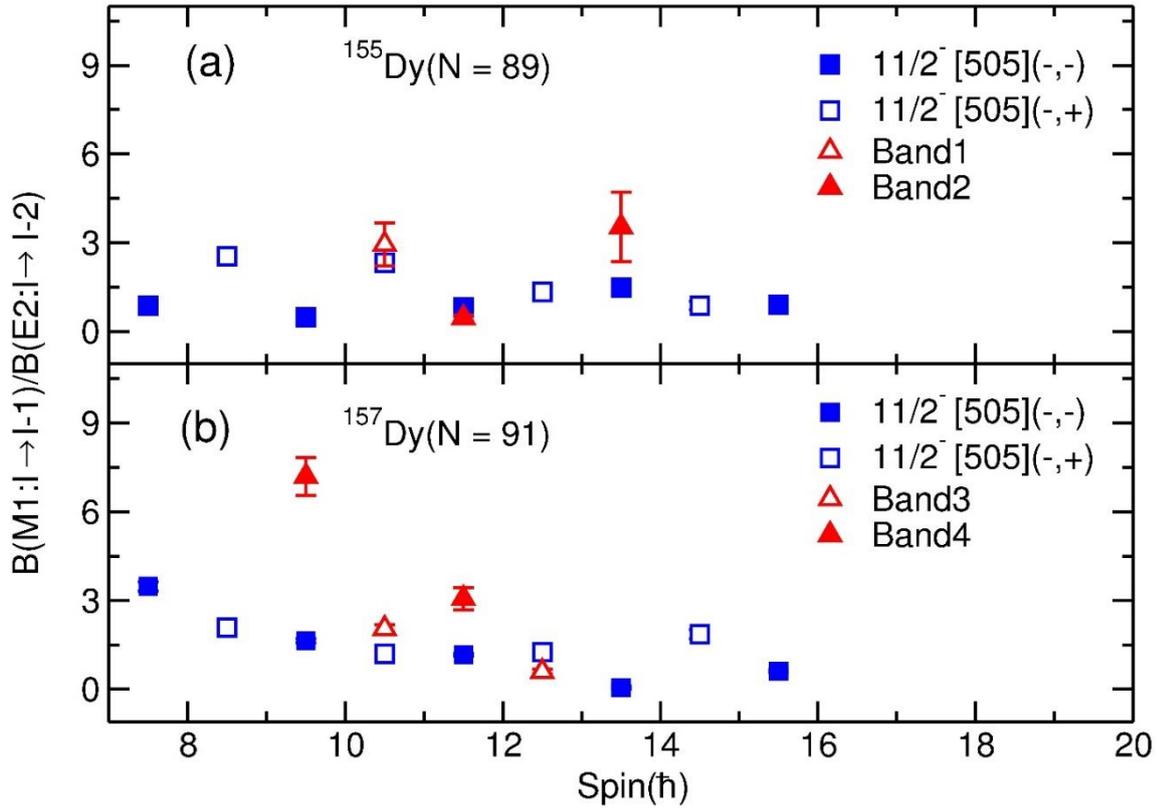

*Figure 7*: (Color online) Experimental B(M1;I →I-1)/B(E2;I →I-2) values for ΔI = 1 M1 intra-band transitions pertaining to the [505]11/2⁻ bands and newly established $K^\pi$ =15/2⁻ bands in (a) $^{157}$Dy and (b) $^{155}$Dy.

## B. *K*= 2⁺ γ vibrational bands built on the [505]11/2⁻ in $^{155}$Dy and $^{157}$Dy

Indeed, these new high-*K* band structures identified in both $^{155}$Dy (i.e., Bands 1 and 2) and $^{157}$Dy (i.e., Bands 3 and 4) share a lot of similarities with the bands that were recently identified as a $K^\pi$ =2⁺ γ vibration, built on the [505]11/2⁻ orbital, in $^{155}$Gd [17, 45]. The decay pattern of these structures is also consistent with those observed for the $K^\pi = 2^+$ γ vibrational bands in the neighbouring even-even nuclei. The energy difference between these structures and their intrinsic states also compare very well with the excitation levels of the γ vibrational bands relative to the ground state bands in the neighbouring even-even-nuclei.

Recently, Wang et al., [18] observed that systematics of band-head energies of the γ-bands in odd-A (odd-N) nuclei are systematically lying very close in energy to those in their neighbouring N+1 even-even nucleus. For instance, the band-head energy of the γ-band is 852-keV relative to the [505]11/2⁻ state in $^{163}$Er, which is close to the 860.3-keV, which is the band-head of the γ-band in $^{164}$Er. The same feature is observed between $^{165}$Er and $^{166}$Er where the γ band-head is at 767-keV relative to the [505]11/2⁻ state, and is close to the 786-keV energy of the lowest-lying $K^\pi=2^+$ band-head in $^{166}$Er. The same situation is observed for the newly established γ-bands conjectured to be $K^\pi=2^+$ γ bands built on the [505]11/2⁻ state in the odd-A Dy isotopes in the A ≈ 150-160 mass region. For instance, the band-head energy for the $K^\pi=15/2^+$ band in $^{155}$Dy is at $E_x$ = 830-keV (see Fig. 1) and is close in energy to the band-head of the first excited $K^\pi=2^+$ band in $^{156}$Dy, at $E_x$ = 890-keV [13]. Similarly, the same feature is observed between the $K^\pi=15/2^+$ band in $^{157}$Dy ($E_x$ = 928-keV) [see Fig. 2] and the $K^\pi=2^+$ γ-band band-head in $^{158}$Dy ($E_x$ = 946-keV) [9]. This trend is also evident in the γ-band that was recently identified in $^{155}$Gd [17]. For instance, its relative band-head energy is 1161-keV and is close to 1154-keV, which is the γ-band band-head in $^{156}$Gd [36]. A systematic review for dysprosium and gadolinium isotopes demonstrating this relation is given in Fig. 8. Clearly, both isotopes demonstrate a similar trend as a function of the neutron number N.

In Fig. 9 we also plot the systematics of the level energies of the γ-bands in the odd-N dysprosium isotopes to demonstrate further similarities with those in the neighbouring even-even isotopes. In fact, the $K^\pi$ = 15/2⁻ γ-bands built on the [505]11/2⁻ orbital seem to mimic those observed in even-even nuclei. In particular, the $K^\pi$ = 15/2⁻ γ-bands run parallel to their intrinsic configuration, which is the [505]11/2⁻ bands. This feature is consistent with the behaviour of the $K^\pi$ = 2⁺ γ vibrational bands in the neighbouring even-even nuclei. In effect, recent studies have demonstrated that γ-bands faithfully track their intrinsic configuration, particularly the odd-spin members [8–13, 33], and this seems to be a general or rather a notable feature for γ-bands in N ≈ 90 region.

In even-even nuclei, the M1 (ΔI = 1) intra-band transitions of the γ-bands are generally known to be very weak (or non-existent) due to the collective nature of γ-bands [5–7, 9, 34]. Indeed, the γ-vibrational bands of the $^{154}$Dy [8] and $^{156}$Dy [9, 13] cores also have very weak (and/or non-existent) ΔI = 1 M1 intra-band transitions. This seems to be a general feature of γ-bands, particularly those of the nuclei in the 150 to 160 mass region [7–9]. The apparently strong M1 intra-band transitions observed for both γ-bands of $^{155}$Dy and $^{157}$Dy can therefore be attributed to coupling of the $K^\pi=2^+$ γ excitations of the even-even cores ($^{154}$Dy and $^{156}$Dy) to the high-$K$ value of the [505]11/2⁻ single−particle orbital itself.

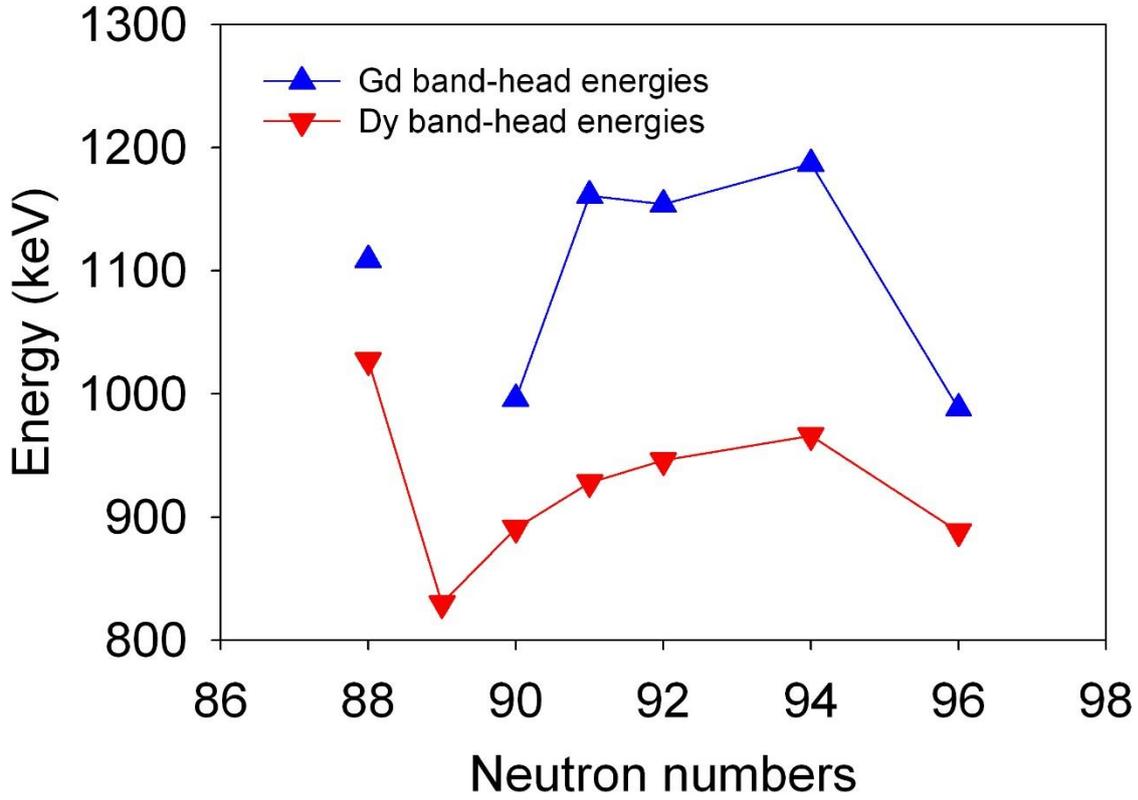

*Figure 8:* (Color online) The systematic review of band-head energies for γ-bands in the even (with respect to the ground state bands) and odd-N Gd and Dy isotopes (from N = 88 to N = 96, i.e. $^{152}$Gd [9], $^{153}$Gd [37], $^{154}$Gd [9, 35], $^{155}$Gd [17], $^{156}$Gd [36], $^{158}$Gd [38], $^{160}$Gd [39], $^{154}$Dy [8], $^{155}$Dy [current work], $^{156}$Dy [13], $^{157}$Dy [current work], $^{158}$Dy [9], $^{160}$Dy [40], $^{162}$Dy [40]). Note that the excitation energy given for the K=15/2⁻ band-heads (in the odd-N nuclei) has been subtracted relative to the excitation energy of the lowest lying state of the K = 11/2⁻ band. The discontinuity in the Gd isotopes can be attributed to the missing band-head (yet to be observed) energy of the γ-band in $^{155}$Gd.

Clearly, in many respects, the $K^\pi = 15/2^-$ γ-bands in $^{155}$Dy and $^{157}$Dy share a lot of structural similarities with those found in their neighbouring even-even nuclei and this therefore confirms the suggestion given above that they represent a coupling of the $K^\pi = 2^+$ γ vibrational bands of the even-even core to the [505]11/2⁻ orbital. With these findings, $^{155}$Dy and $^{157}$Dy are the first odd-N dysprosium isotopes in which γ vibrational bands built on the [505]11/2⁻ orbital have been observed. In spite of this, it worth mentioning that the $K_<=7/2^-$ band, resulting from the antiparallel coupling of the [505]11/2⁻ orbital with the $K^\pi = 2^+$ vibrational band was not observed in both $^{155}$Dy and $^{157}$Dy.

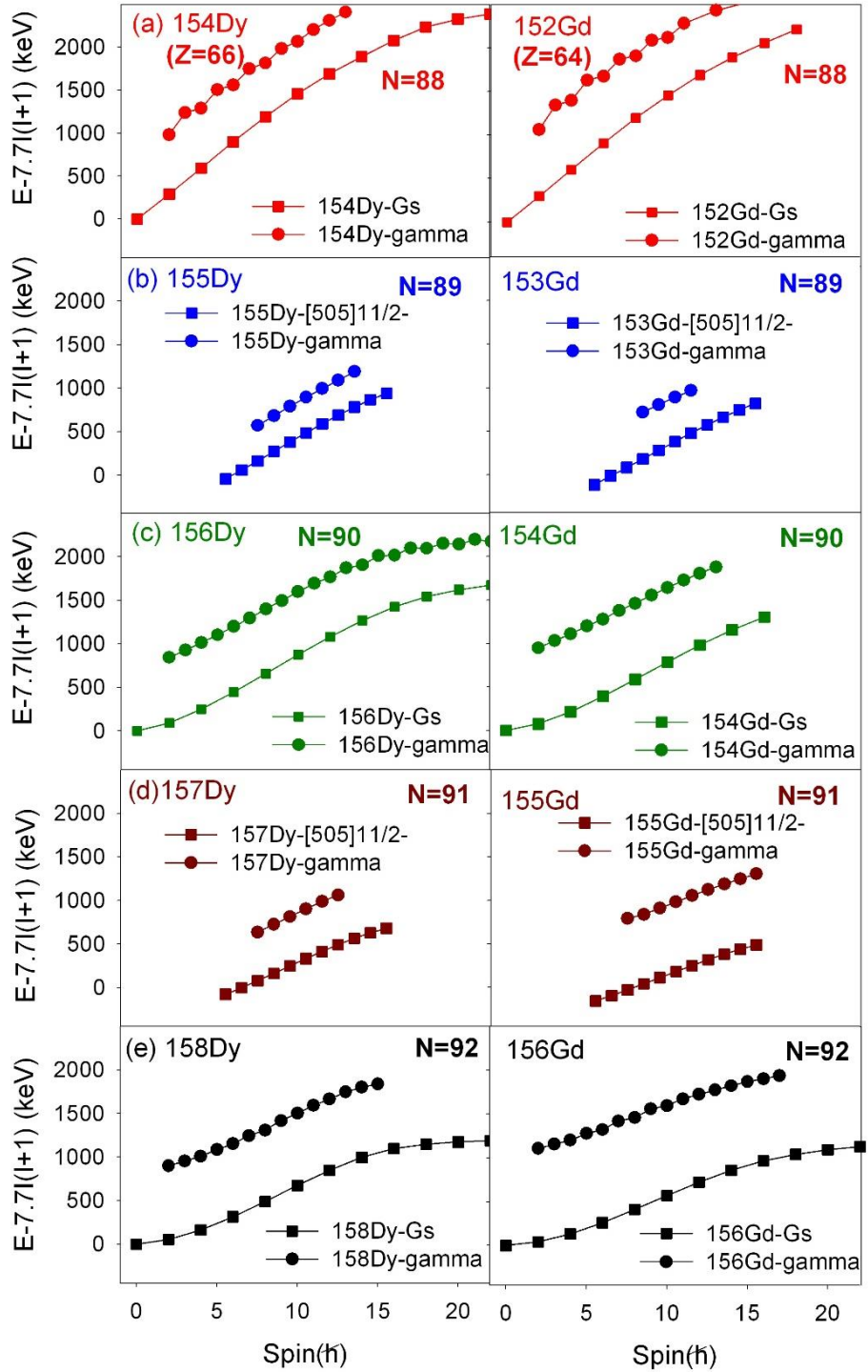

***Fig.9:*** *(Color online) A plot of excitation energy minus the rigid rotor energy, showing the systematics of γ-bands with respect to their intrinsic states in even-even and odd-A isotopes of gadolinium and dysprosium (from N = 88 to N = 92 i.e., $^{152}$Gd [9], $^{153}$Gd [37], $^{154}$Gd [9, 35], $^{155}$Gd [17], $^{156}$Gd [36], $^{154}$Dy [8], $^{155}$Dy [current work], $^{156}$Dy [13], $^{157}$Dy [current work] and $^{158}$Dy [9]).*

## C. Elusive coupling of the [505]11/2⁻ neutron orbital to the first excited $K=0^+$ core excitations of $^{154,156}$Dy in $^{155,157}$Dy?

As stated above, in odd-A nuclei such as $^{155,157}$Dy isotopes, the single nucleons in a Nilsson orbital [N,n$_x$,Λ]Ω are expected to couple to the collective states of the even-even core. And indeed, the present study has successfully identified for the first time, the $K_>=15/2^-$ band, resulting from a parallel coupling of the [505]11/2⁻ orbital with a $K^\pi = 2^+$ vibrational band in both $^{155}$Dy and $^{157}$Dy. In spite of this, the respective coupling of the first excited $0^+$ states or so-called β vibrational bands at 661- and 676-keV in $^{154}$Dy and $^{156}$Dy to the [505]11/2⁻ orbital, to produce a $K=11/2^-$ band, was not observed in both $^{155}$Dy and $^{157}$Dy. It is worth noting that an equivalent situation has also been observed by Sharpey-Schafer et al., in the neighbouring $^{155}$Gd [17, 45].

According to Refs. [17, 35, 45], the first excited $0_2^+$ states in the N ≈ 90 nuclei are actually not collective β-vibrations, but have properties resembling pairing isomers [46]. The latter refers to states which are lowered into the pairing gap by a configuration dependent pairing interaction [13, 17, 35, 45]. In such a scenario, the $0_2^+$ bands can be understood as neutron 2p-2h seniority-zero states formed by raising two neutrons out of the ground-state configuration into the upsloping $h_{11/2}$ neutron orbit. This up-sloping orbit is raised from the lower filled shell to the Fermi surface at the onset of deformation after the N = 82 shell closure. Due to quadrupole pairing effects (or configuration-dependent pairing interaction), pairs of neutrons in this high-$K$ "oblate" polar orbit do not take part in the normal monopole pairing produced by the high density of low-Ω "prolate" equatorial rotationally aligned orbitals [13]. And due to the small overlap of the wave-functions of prolate and oblate single-particle states, [505]² configurations are decoupled from the pairing produced by the interactions of low-Ω $i_{13/2}$ neutron pairs that cause back-bending in the rotational bands of the deformed nuclei near N = 90 [13, 47]. Consequently, the weakening of the monopole pairing strength and the increase of the quadrupole pairing strength could bring 2p-2h $0^+$ states within the pairing gap of 2Δ. This leaves a pair of neutrons in the $h_{11/2}$[505] orbital as "spectators" in the wider pairing correlations of the nucleus, which then forms its own low-lying $0^+$ state, well within the pairing gap, which in turn can compete with β bands [11].

Furthermore, according to Refs. [17, 35, 45], the conspicuous absence of the coupling between $K^{\pi} = 0^+$ states and the [505]11/2$^-$ orbital in $^{155}$Gd as well as the rest of the neighbouring odd-N nuclei (such as $^{155}$Dy and $^{157}$Dy) is mainly because the underlying configuration of the first-excited $K^{\pi} = 0^+$ states, in even-even nuclei in the transitional N ≈ 90 rare-earth region, are supposedly comprised of essentially an $h_{11/2}$ pair of neutrons. Thus, the coupling of the $K^{\pi} = 0^+$ band with 11/2$^-$[505] orbital in the neighbouring odd neutron nucleus should be Pauli-blocked. Conversely, the coupling of the $K=2^+$ γ-band with the [505]11/2$^-$ orbital is expected to be observed (as is the case in $^{155}$Dy and $^{157}$Dy), since it is not Pauli-blocked.

The consideration of the new ($^{155}$Dy and $^{157}$Dy) and existing ($^{155}$Gd) evidence supports the interpretation that the first excited $0^+$ states are due to configuration-dependent pairing interaction as conjectured by Refs. [17, 35]. In effect, the apparent absence of the coupling of the $h_{11/2}$ neutron orbitals to the first-excited $K^{\pi} = 0^+$ states observed in the current work is consistent with the suggestion that the first excited $0^+$ states in the N ≈ 90 region are not due to β vibrations but are essentially comprised of a pair of $h_{11/2}$ neutrons [17, 35, 45]. Nevertheless, it is worth noting that this condition cannot be considered as an unambiguous proof since the absence does not necessarily imply the non-existence of these elusive states. Clearly, more evidence is needed before a final conclusion can be reached. Therefore, future experimental and theoretical studies are essential in order to positively identify whether the $h_{11/2}$ neutron orbital plays any role in the absence of the coupling between the $K^{\pi} = 0^+$ states and the [505]11/2$^-$ neutron in odd-A Gd and Dy isotopes in the N ≈ 90 region.

## V. Conclusion

Rotational structures have been measured using the Jurogam II and GAMMASPHERE arrays at low spin following the $^{155}$Gd(α,2n)$^{157}$Dy and $^{148}$Nd($^{12}$C, 3n)$^{157}$Dy reactions at 25 and 65 MeV, respectively. This study has identified for the first time, the $K_>$=15/2$^-$ band, resulting from a parallel coupling of the [505]11/2$^-$ orbital with a $K^{\pi} = 2^+$ γ-band excitations in both $^{155}$Dy and $^{157}$Dy. Just like the $K^{\pi} = 2^+$ excitations observed in the neighbouring even-even nuclei, the γ-bands in both $^{155}$Dy and $^{157}$Dy appear to faithfully track their intrinsic configuration, as a function of spin. With these findings, $^{155}$Dy and $^{157}$Dy are the first odd-N dysprosium isotopes in which γ vibrational bands built on the [505]11/2$^-$ orbital have been observed.

The coupling of the first excited $K^\pi=0^+$ states or the so-called β vibrational bands at 661 and 676 keV in $^{154}$Dy and $^{156}$Dy to the [505]11/2$^-$ orbital, to produce $K^\pi=11/2^-$ bands, was not observed in either $^{155}$Dy and $^{157}$Dy, respectively. These findings, together with the existing evidence presented in Refs. [17, 45], favour the suggestion that the first excited $0^+$ states are due to configuration-dependent pairing interaction formed by the upsloping $h_{11/2}$ neutron orbit around N ≈ 90. However, we are of the view that the apparent absence of that coupling is indeed an essential but not a sufficient condition to prove that the microscopic identity of the first excited $0^+$ state is comprised of essentially an $h_{11/2}$ pair of neutrons as prescribed by Refs. [17, 35, 45]. In light of this, additional theoretical and experimental spectroscopic studies are needed in order to obtain further insight on the nature of the first excited $K^\pi = 0^+$ in even-even nuclei (in the N ≈ 90 region), notwithstanding their elusive coupling to the $h_{11/2}$ neutron orbitals in the neighbouring odd-N nucleus such as $^{155, 157}$Dy isotopes.

## ACKNOWLEDGEMENTS

This work is supported by the South African National Research Foundation under grants (No. 96829, 109711, and 106012). Support for L.B. and P.E.G. was provided by the Natural Sciences and Engineering Research Council of Canada. VW wishes to acknowledge support provided by the German Federal Ministry for education and Research under grant No. 05P19RDFN1. Support was also provided by the U.S. National Science Foundation under Grant Nos. PHY-1203100 (USNA), PHY-1502092 (USNA) and PHY–0754674 (FSU) as well as by the U.S. Department of energy, office of Nuclear of Nuclear Physics, under Contract Nos. DE-AC02-06CH11357 (ANL) and DE-FG02-91ER-40609 (Yale). The authors also acknowledge the support of GAMMAPOOL for the loan of the JUROGAM detectors. JYFL research is supported by the Academy of Finland under the Finnish Centre of Excellence Programme 2006–2011, Contract No. 213503. This work was also partially supported by the National Research, Development and Innovation Fund of Hungary, financed under the K18 funding scheme with project no. K128947, as well as by the European Regional Development Fund (Contract No. GINOP–2.3.3–15–2016–00034).

*Table 3: Experimentally determined properties for the nucleus $^{155}$Dy. This includes excitation levels $E_x$ (in keV), γ-ray energies $E_γ$ (in keV) §, γ-ray intensities $I_γ$, assigned multipolarities (Mult.), spins for the initial $I_i^π$ and final $I_f^π$ states, level in the band of destination Band$_f$, DCO ratios and polarization measurements ($A_p$). All DCO ratios were deduced by gating on stretched E2 transitions. NB//: Quantities within parenthesis are quoted as tentative while empty entries refer to information that could not be obtained.*

| $E_x$ | $E_γ$ | $I_γ$ | Mult | $I_i^π$ | $I_f^π$ | Band$_f$ | DCO | $A_p$ |
|---|---|---|---|---|---|---|---|---|
| \[505\]11/2$^-$ (-, +1/2)$_1$ | | | | | | | | |
| 436 | 201.8 | 10.1(12) | M1/E2 | 13/2$^-$ | 11/2$^-$ | (-, -1/2)$_1$ | | |
| 896 | 238.4 | 4.24(18) | M1/E2 | 17/2$^-$ | 15/2$^-$ | (-, -1/2)$_1$ | | |
| 896 | 459.6 | 1.75(12) | E2 | 17/2$^-$ | 13/2$^-$ | (-, +1/2)$_1$ | | |
| 1418 | 268.2 | 2.96(19) | M1/E2 | 21/2$^-$ | 19/2$^-$ | (-, -1/2)$_1$ | | |
| 1418 | 522.3 | 1.78(13) | E2 | 21/2$^-$ | 17/2$^-$ | (-, +1/2)$_1$ | | |
| 1990 | 291.3 | 1.22(9) | M1/E2 | 25/2$^-$ | 23/2$^-$ | (-, -1/2)$_1$ | | |
| 1990 | 571.9 | 1.57(16) | E2 | 25/2$^-$ | 21/2$^-$ | (-, +1/2)$_1$ | | |
| 2598 | 307.5 | 0.92(10) | M1/E2 | 29/2$^-$ | 27/2$^-$ | (-, -1/2)$_1$ | | |
| 2598 | 608.1 | 2.07(25) | E2 | 29/2$^-$ | 25/2$^-$ | (-, +1/2)$_1$ | | |
| \[505\]11/2$^-$ (-, -1/2)$_1$ | | | | | | | | |
| 657 | 221.4 | 4.42(17) | M1/E2 | 15/2$^-$ | 13/2$^-$ | (-, +1/2)$_1$ | | |
| 657 | 422.9 | 4.4(3) | E2 | 15/2$^-$ | 11/2$^-$ | (-, -1/2)$_1$ | | |
| 1150 | 254.1 | 1.43(12) | M1/E2 | 19/2$^-$ | 17/2$^-$ | (-, +1/2)$_1$ | | |
| 1150 | 492.6 | 3.7(3) | E2 | 19/2$^-$ | 15/2$^-$ | (-, -1/2)$_1$ | | |
| 1699 | 280.6 | 1.73(11) | M1/E2 | 23/2$^-$ | 21/2$^-$ | (-, +1/2)$_1$ | | |
| 1699 | 548.8 | 3.39(20) | E2 | 23/2$^-$ | 19/2$^-$ | (-, -1/2)$_1$ | | |
| 2291 | 300.6 | 1.38(12) | M1/E2 | 27/2$^-$ | 25/2$^-$ | (-, +1/2)$_1$ | | |

| $E_x$ | $E_\gamma$ | $I_\gamma$ | Mult | $I_i^\pi$ | $I_f^\pi$ | Band$_f$ | DCO | $A_p$ |
|---|---|---|---|---|---|---|---|---|
| \multicolumn{9}{c}{[505]11/2$^-$ (-, -1/2)$_1$ *continues*} | | | | | | | | |
| 2291 | 591.9 | 1.74(16) | E2 | 27/2$^-$ | 23/2$^-$ | (-, -1/2)$_1$ | | |
| 2909 | 310.9 | 0.89(11) | M1/E2 | 31/2$^-$ | 29/2$^-$ | (-, +1/2)$_1$ | | |
| 2909 | 618.4 | 2.05(17) | E2 | 31/2$^-$ | 27/2$^-$ | (-, -1/2)$_1$ | | |
| \multicolumn{9}{c}{*Band1*} | | | | | | | | |
| 1305 | 241.4 | 0.25(6) | M1/E2 | (17/2$^-$) | (15/2$^-$) | Band2 | | |
| 1305 | 409.8 | 0.52(12) | M1/E2 | (17/2$^-$) | 17/2$^-$ | (-, +1/2)$_1$ | | |
| 1305 | 648.3 | 0.69(14) | M1/E2 | (17/2$^-$) | 15/2$^-$ | (-, -1/2)$_1$ | | |
| 1305 | 869.4 | 0.48(10) | E2 | (17/2$^-$) | 13/2$^-$ | (-, +1/2)$_1$ | | |
| 1827 | 267.0 | 2.0(4) | M1/E2 | (21/2$^-$) | (19/2$^-$) | Band2 | | |
| 1827 | 409.0 | 0.05(1) | M1/E2 | (21/2$^-$) | 21/2$^-$ | (-, +1/2)$_1$ | | |
| 1827 | 521.5 | 0.97(16) | E2 | (21/2$^-$) | (17/2$^-$) | Band1 | | |
| 1827 | 677.2 | 0.05(1) | M1/E2 | (21/2$^-$) | 19/2$^-$ | (-, -1/2)$_1$ | | |
| 1827 | 931.3 | 0.25(8) | E2 | (21/2$^-$) | 17/2$^-$ | (-, +1/2)$_1$ | | |
| 2390 | 285.6 | 0.17(5) | M1/E2 | (25/2$^-$) | (23/2$^-$) | Band2 | | |
| 2390 | 563.0 | 0.18(2) | E2 | (25/2$^-$) | (21/2$^-$) | Band1 | | |
| 2963 | 573.0 | 0.68(16) | E2 | (29/2$^-$) | (25/2$^-$) | Band1 | | |
| \multicolumn{9}{c}{*Band2*} | | | | | | | | |
| 1064 | 627.0 | 0.31(20) | M1/E2 | (15/2$^-$) | 13/2$^-$ | (-, +1/2)$_1$ | | |
| 1064 | 829.8 | 0.15(12) | E2 | (15/2$^-$) | 11/2$^-$ | (-, -1/2)$_1$ | 1.09(19) | |
| 1560 | 254.5 | 2.39(19) | M1/E2 | (19/2$^-$) | (17/2$^-$) | Band1 | | |
| 1560 | 410.2 | 0.52(16) | M1/E2 | (19/2$^-$) | 19/2$^-$ | (-, -1/2)$_1$ | | |
| 1560 | 495.9 | 0.18(12) | E2 | (19/2$^-$) | (15/2$^-$) | Band2 | | |

| $E_x$ | $E_\gamma$ | $I_\gamma$ | Mult | $I_i^\pi$ | $I_f^\pi$ | Band$_f$ | DCO | $A_p$ |
|---|---|---|---|---|---|---|---|---|
| | | | *Band2 continues* | | | | | |
| 1560 | 664.3 | 0.47(12) | M1/E2 | (19/2$^-$) | 17/2$^-$ | (-, +1/2)$_1$ | | |
| 1560 | 902.8 | 0.33(10) | E2 | (19/2$^-$) | 15/2$^-$ | (-, -1/2)$_1$ | | |
| 2104 | 277.4 | 0.46(8) | M1/E2 | (23/2$^-$) | (21/2$^-$) | Band1 | | |
| 2104 | 544.4 | 1.55(15) | E2 | (23/2$^-$) | (19/2$^-$) | Band2 | | |
| 2699 | (308.9) | 1.03(13) | M1/E2 | (27/2$^-$) | (25/2$^-$) | Band1 | | |
| 2699 | 594.5 | 0.51(16) | E2 | (27/2$^-$) | (23/2$^-$) | Band2 | | |

$^{§1}$Typical uncertainty of 0.3 keV and up to 0.5 keV for weak transitions ($I_\gamma<1$).

**Table 4:** Experimentally determined properties for the nucleus $^{157}$Dy. This includes excitation levels $E_x$ (in keV), γ-ray energies $E_γ$ (in keV) §, γ-ray intensities $I_γ$, assigned multipolarities (Mult.), spins for the initial $I_i^π$ and final $I_f^π$ states, level in the band of destination Band$_f$, DCO ratios and polarization measurements ($A_p$). All DCO ratios were deduced by gating on stretched E2 transitions. NB//: Quantities within parenthesis are quoted as tentative while empty entries refer to information that could not be obtained.

| $E_x$ | $E_γ$ | $I_γ$ | Mult | $I_i^π$ | $I_f^π$ | Band$_f$ | DCO | $A_p$ |
|---|---|---|---|---|---|---|---|---|
| \multicolumn{9}{c}{[505]11/2$^-$ (-, +1/2)$_1$} |
| 375 | 175.5 | 1.58(16) | M1/E2 | 13/2$^-$ | 11/2$^-$ | (-, -1/2)$_1$ |  | -0.04(2) |
| 785 | 214.3 | 27.8(8) | M1/E2 | 17/2$^-$ | 15/2$^-$ | (-, -1/2)$_1$ | 0.38(3) | -0.03(1) |
| 785 | 410.4 | 10.9(3) | E2 | 17/2$^-$ | 13/2$^-$ | (-, +1/2)$_1$ | 1.27(12) | 0.08(2) |
| 1263 | 246.4 | 7.12(22) | M1/E2 | 21/2$^-$ | 19/2$^-$ | (-, -1/2)$_1$ |  | -0.03 (2) |
| 1263 | 477.7 | 6.80(21) | E2 | 21/2$^-$ | 17/2$^-$ | (-, +1/2)$_1$ | 0.99(13) | 0.08(2) |
| 1793 | 270.6 | 1.92(6) | M1/E2 | 25/2$^-$ | 23/2$^-$ | (-, -1/2)$_1$ |  | -0.02(1) |
| 1793 | 529.9 | 2.22(7) | E2 | 25/2$^-$ | 21/2$^-$ | (-, +1/2)$_1$ | 0.99(17) | 0.09(2) |
| 2359 | 286.8 | 0.29(2) | M1/E2 | 29/2$^-$ | 27/2$^-$ | (-, -1/2)$_1$ |  | -0.41(4) |
| 2359 | 566.6 | 0.84(3) | E2 | 29/2$^-$ | 25/2$^-$ | (-, +1/2)$_1$ |  |  |
| \multicolumn{9}{c}{[505]11/2$^-$ (-, -1/2)$_1$} |
| 571 | 196.1 | 48.5(20) | M1/E2 | 15/2$^-$ | 13/2$^-$ | (-, +1/2)$_1$ |  | -0.02(1) |
| 571 | 371.6 | 9.1(3) | E2 | 15/2$^-$ | 11/2$^-$ | (-, -1/2)$_1$ | 1.02(12) | 0.10(3) |
| 1017 | 231.3 | 14.2(4) | M1/E2 | 19/2$^-$ | 17/2$^-$ | (-, +1/2)$_1$ | 0.32(16) | -0.05(1) |
| 1017 | 445.6 | 8.5(3) | E2 | 19/2$^-$ | 15/2$^-$ | (-, -1/2)$_1$ | 1.09(19) | 0.07(2) |
| 1522 | 259.3 | 3.57(11) | M1/E2 | 23/2$^-$ | 21/2$^-$ | (-, +1/2)$_1$ | 0.32(7) | -0.03(2) |
| 1522 | 505.7 | 4.03(12) | E2 | 23/2$^-$ | 19/2$^-$ | (-, -1/2)$_1$ | 0.96(17) | 0.05(4) |
| 2073 | 279.8 | 0.74(3) | M1/E2 | 27/2$^-$ | 25/2$^-$ | (-, +1/2)$_1$ | 0.43(4) | -0.23(3) |
| 2073 | 550.4 | 1.14(4) | E2 | 27/2$^-$ | 23/2$^-$ | (-, -1/2)$_1$ | 0.99(13) | 0.10(7) |

| $E_x$ | $E_\gamma$ | $I_\gamma$ | Mult | $I_i^\pi$ | $I_f^\pi$ | Band$_f$ | DCO | $A_p$ |
|---|---|---|---|---|---|---|---|---|
| \[505\]11/2$^-$ (-, -1/2)$_1$ *continues* | | | | | | | | |
| 2651 | 291.9 | 0.31(2) | M1/E2 | 31/2$^-$ | 29/2$^-$ | (-, +1/2)$_1$ | | |
| 2651 | 578.7 | 0.31(2) | E2 | 31/2$^-$ | 27/2$^-$ | (-, -1/2)$_1$ | | |
| *Band3* | | | | | | | | |
| 1347 | 220.1 | 1.35(5) | M1/E2 | 17/2$^-$ | 15/2$^-$ | Band4 | | |
| 1347 | 562.0 | 1.57(6) | M1/E2 | 17/2$^-$ | 17/2$^-$ | (-, +1/2)$_1$ | 0.69(4) | |
| 1347 | 776.3 | 2.34(8) | M1/E2 | 17/2$^-$ | 15/2$^-$ | (-, -1/2)$_1$ | 0.57(8) | -0.02(1) |
| 1347 | 972.4 | 2.46(8) | E2 | 17/2$^-$ | 13/2$^-$ | (-, +1/2)$_1$ | 1.16(10) | 0.05(4) |
| 1834 | 250.7 | 1.07(4) | M1/E2 | 21/2$^-$ | 19/2$^-$ | Band4 | | |
| 1834 | 486.7 | 0.62(3) | E2 | 21/2$^-$ | 17/2$^-$ | Band3 | | |
| 1834 | 817.4 | 0.57(3) | M1/E2 | 21/2$^-$ | 19/2$^-$ | (-, -1/2)$_1$ | | -0.03(2) |
| 1834 | 1048.7 | 2.01(7) | E2 | 21/2$^-$ | 17/2$^-$ | (-, +1/2)$_1$ | | |
| 2363 | 268.2 | 0.18(2) | M1/E2 | (25/2$^-$) | (23/2$^-$) | Band4 | | |
| 2363 | 528.8 | 0.45(3) | E2 | (25/2$^-$) | 21/2$^-$ | Band3 | | |
| 2363 | 1099.8 | 0.43(2) | E2 | (25/2$^-$) | 21/2$^-$ | (-, +1/2)$_1$ | | |
| *Band4* | | | | | | | | |
| 1127 | 752.3 | 1.53(7) | M1/E2 | 15/2$^-$ | 13/2$^-$ | (-, +1/2)$_1$ | | -0.96(2) |
| 1127 | 927.8 | 2.70(14) | E2 | 15/2$^-$ | 11/2$^-$ | (-, -1/2)$_1$ | 1.11(9) | 0.03(1) |
| 1583 | 236.0 | 1.25(5) | M1/E2 | 19/2$^-$ | 17/2$^-$ | Band3 | | -0.02(1) |
| 1583 | 456.0 | 0.18(2) | E2 | 19/2$^-$ | 15/2$^-$ | Band4 | | |
| 1583 | 567.1 | 0.71(4) | M1/E2 | 19/2$^-$ | 19/2$^-$ | (-, -1/2)$_1$ | | |
| 1583 | 798.0 | 1.35(5) | M1/E2 | 19/2$^-$ | 17/2$^-$ | (-, +1/2)$_1$ | | |
| 1583 | 1012.4 | 2.50(9) | E2 | 19/2$^-$ | 15/2$^-$ | (-, -1/2)$_1$ | | 0.08(1) |
| 2095 | 260.6 | 0.36(3) | M1/E2 | (23/2$^-$) | 21/2$^-$ | Band3 | | |

| $E_x$ | $E_\gamma$ | $I_\gamma$ | Mult | $I_i^\pi$ | $I_f^\pi$ | Band$_f$ | DCO | $A_p$ |
|---|---|---|---|---|---|---|---|---|
| | | | *Band4 continues* | | | | | |
| 2095 | 511.3 | 0.16(2) | E2 | (23/2⁻) | 19/2⁻ | Band4 | | |
| 2095 | 1078.0 | 0.57(3) | E2 | (23/2⁻) | 19/2⁻ | (-, -1/2)₁ | | |

[§1] Typical uncertainty of 0.3 keV and up to 0.5 keV for weak transitions ($I_\gamma<1$).